\documentclass[12pt]{article}
\usepackage{amsmath,amssymb,amsfonts,epsfig,graphicx,euscript}%
\usepackage{color}

\newcommand{\bse}{\begin{subequations}}
\newcommand{\ese}{\end{subequations}}
\newcommand{\be}{\begin{equation}}
\newcommand{\ee}{\end{equation}}
\newcommand{\bea}{\begin{eqnarray}}
\newcommand{\eea}{\end{eqnarray}}
\newcommand{\ba}{\begin{array}}
\newcommand{\ea}{\end{array}}

\def\a{\alpha}

\def\N{{\mathcal{N}}}

\begin{document}
\hfill%
\vspace{1cm}
\begin{center}
{ \Large{\textbf{$\alpha'$-Corrected Chiral Magnetic Effect}}} 
\vspace*{2cm}
\begin{center}
{\bf M. Ali-Akbari\footnote{aliakbari@theory.ipm.ac.ir}, S. F. Taghavi\footnote{s.f.taghavi@ipm.ir}}\\%
\vspace*{0.4cm}
{\it {School of Particles and Accelerators,\\ Institute for Research in Fundamental Sciences (IPM),\\
P.O.Box 19395-5531, Tehran, Iran}}  \\

\vspace*{1.5cm}
\end{center}
\end{center}

\vspace{.5cm}
\bigskip
\begin{center}
\textbf{Abstract}
\end{center}
Using the AdS/CFT correspondence, the effect of $\alpha'$-correction
on the value of Chiral Magnetic Effect(CME) is computed by adding a
number of spinning probe D7-branes in the $\alpha'$-corrected
background. We numerically show that the magnitude of CME rises in
the presence of $\alpha'$-correction for massive solutions and this
increase is more sensible at higher temperatures. However, this
value does not change for massless solution. Although some of the
D7-brane embeddings have no CME, after applying the
$\alpha'$-correction they find a non-zero value for the CME. We also
show that the effect of $\alpha'$-correction removes the singularity
from some of the D7-brane embeddings.

\newpage

\tableofcontents

\section{Introduction}

A new phase of Quantum Chromodynamics (QCD) called Quark-Gluon
plasma (QGP) is produced at Relativistic Heavy Ion Collider (RHIC)
and Large Hadron Collider (LHC) by colliding two pancakes of heavy
nuclei such as Gold or Lead at a relativistic speed. It was
realized that the QGP is a strongly coupled perfect fluid with very low
viscosity over entropy density, $\eta/s$. Therefore in order to
describe it the perturbative methods are inapplicable. Different
properties of the QGP such as rapid thermalization, elliptic flow,
jet quenching parameter and quarkonium dissociation have been considerably
studied \cite{Ollitrault,Snellings,CasalderreySolana}.

One of the interesting properties of QGP is the Chiral Magnetic Effect (CME). The presence of
a strong magnetic field in the very early stages of heavy ion
collision and its accompanying non-trivial gluon field
configurations lead to the CME, which is the generation of an
electric current along the strong magnetic field
\cite{Kharzeev:2004ey, Kharzeev:2007tn, Kharzeev:2007jp,
Fukushima:2008xe,Kharzeev:2009fn}. In free massless QCD the CME can
be described in the following way. This theory enjoys two global
$U(1)$ symmetries: vector symmetry, $U(1)_V$,  and axial symmetry,
$U(1)_A$. If we assume that the global vector symmetry is preserved at
quantum level, $U(1)_A$ will be anomalous. The axial charge
associated to this symmetry, given by the difference between the
number of fermions with left-handed and right-handed helicities, is
proportional to the winding number of non-trivial gauge field
provided that the left-handed and right-handed fermions are
initially equal. The spin of fermions is tightly aligned along the
strong magnetic field. For a non-zero winding number, in order to
have a non-zero axial charge the momentum direction of some fermions
depending on the sign of winding number must be altered. This
phenomena leads to an electric current along the magnetic field, the
CME. There are various ways to calculate this current in the
literature \cite{Kharzeev:2007jp, Fukushima:2008xe,Kharzeev:2009fn}.

Consider an effective Lagrangian which includes the original QCD
Lagrangian and an extra term $\theta(t,x) \tilde{F}^{\mu\nu}
F_{\mu\nu}$ where the form of the extra term is fixed by the anomaly and
$\theta(t,x)$ is a non-dynamical axion field \cite{Kharzeev:2009fn}.
Assuming a space-independent $\theta(t,x)$, which means
$\theta(x,t)\equiv\theta(t)$, the time derivative of $\theta(t)$ can
be identified by $\mu_5$ which is the axial chemical potential
\cite{Kharzeev:2009fn}. Using the effective Lagrangian the value of
electric
current is computed by \cite{Kharzeev:2009fn} %
\be %
 J=\frac{\mu_5}{2\pi^2}B~. %
\ee %
Since the QGP is a strongly coupled fluid, AdS/CFT correspondence
seems to be a suitable framework to describe the CME \cite{Rebhan}.

The AdS/CFT correspondence \cite{ads/cft} states that type IIB
string theory on $AdS_5\times S^5$ geometry, describing the near
horizon geometry of a stack of $N_c$ extremal D3-branes, is dual to
the four-dimensional ${\cal{N}}=4$ super Yang-Milles (SYM) theory
with gauge group $SU(N_c)$. One of the most interesting properties
of this correspondence is that it is a strong/weak correspondence. In
particular in the large $N_c$ limit a strongly coupled SYM theory is
dual to the type IIb supergravity which provides a useful tool
to study the strongly coupled regime of the SYM theory. This
correspondence also generalized to the thermal SYM. As a result
a thermal SYM theory corresponds to the supergravity in an
AdS-Schwarzschild background where SYM theory temperature is
identified with the Hawking temperature of AdS black hole
\cite{Witten}.

In the context of AdS/CFT correspondence matter fields in the
fundamental representation can be studied by introducing space
filling flavor D7-branes in the probe limit \cite{Karch} which means
that the number of flavor D7-branes, $N_f$, are very much smaller
than the number of D3-branes. The open strings stretched between
D3-branes and flavor D7-branes give rise to $\N=2$ hypermultiplet in
the gauge theory side. Mass of the hypermultiplet is proportional to
the separation of D3- and D7-branes in transverse directions. This
system, D3-D7 system, is a famous candidate to describe QCD-like
theories \cite{Erdmenger}. Different aspects of adding flavor
D7-branes in different backgrounds have been studied in the
literature. Here we are interested in introducing D7-branes in the
$\alpha'$-corrected background to study the effect of higher order
correction on the CME.

The $\alpha'$-correction has been studied
in \cite{AliAkbari,aliakbari9,Aliakbari10,aliakbari11,AliAkbari1,AliAkbari2,AliAkbari3}. The main motivation to
consider $\alpha'$-correction comes from the fact that string theory
contains higher derivative corrections arising from stringy effects.
The leading order correction in $1/\lambda$, where $\lambda$ is the
t'Hooft coupling constant in gauge theory side, arises from stringy
corrections to the low energy effective action of type IIb supergravity,
$\alpha'^3{\cal{R}}^4$. An understanding of how these computations are
affected by finite $\lambda$ correction is essential for precise
theoretical predictions.

\section{Review on Holographic CME}

Using D3-D7 system an interesting holographic description for the
CME is introduced in \cite{CME}. Our aim in this section is to give
a short review on this paper. In order to do that, we consider a supersymmetric
intersection of $N_c$ D3-branes and $N_f$ probe
D7-branes as 
\be %
\begin{array}{ccccccccccc}
                   & 0 & 1 & 2 & 3 & 4 & 5 & 6 & 7 & 8 & 9 \\
                  D3 & \times & \times & \times & \times &  &  &  &  &  &  \\
                  D7 & \times & \times & \times & \times & \times & \times & \times & \times &  &
\end{array}
\ee %
where the probe limit means that $\frac{N_f}{N_c}\ll1$. This
configuration holographicly describes a ${\cal{N}}=4$ superconformal
$SU(N_c)$ Yang-Mills theory coupled to ${\cal{N}}=2$ hypermultiplet
in the fundamental representation of the gauge group. $SO(6)$
symmetry in the transverse directions to the D3-branes corresponds
to the R-symmetry of ${\cal{N}}=4$ SYM. D7-branes break this
symmetry to $SO(4)\times U(1)$ corresponding to the rotational
symmetries in 4567 directions and 89-plane. The rotational symmetry
in 89-plane which is dual to the $U(1)_R$ subgroup of ${\cal{N}}=2$
R-symmetry, $SU(2)_R\times U(1)_R$, is identified by the axial
symmetry, $U(1)_A$. In QCD, $U(1)_A$ is anomalous and as we will see
this anomaly in the gravity side is derived from the Wess-Zumino (WZ)
part of D7-brane action.

The mass of the ${\cal{N}}=2$ hypermultiplet is considered as a distance
between D3- and D7-branes in the 89-plane. If this distance is zero the
matter fields in fundamental representation are massless. Otherwise,
one can define a complex mass, $|m|e^{i\varphi}$, where $|m|$ is
proportional to the distance between D3- and D7-branes.

In the large $N_c$ and large t'Hooft coupling limit,
$\lambda=g_{YM}^2N_c=4\pi g_s N_c$, the D3-branes are replaced by $AdS_5\times
S^5$ background which is dual to strongly coupled ${\cal{N}}=4$ SYM
at zero temperature. At finite temperature they are replaced by
$AdS$-Schwarzschild black hole where the Hawking temperature of $AdS$
black hole is identified with the temperature of strongly
coupled ${\cal{N}}=4$ SYM. The $AdS$-Schwarzschild metric  in units of the $AdS_5$ radius is%
\be\begin{split}\label{background} %
 ds^2&=-|g_{tt}|dt^2+g_{xx}d\vec{x}^2+g_{rr}dr^2+g_{RR}dR^2+g_{ss}ds_{S^3}^2+g_{\varphi\varphi}d{\varphi}^2, \cr %
  g_{rr}&=g_{RR}=\frac{1}{\rho^2},\ \  g_{ss}=\frac{r^2}{\rho^2},\ \ g_{\varphi\varphi}=\frac{1}{\rho^2}, \cr %
  |g_{tt}|&=\rho^2\frac{\gamma^2}{2}\frac{f^2(\rho)}{H(\rho)},\ \
  g_{xx}=\rho^2\frac{\gamma^2}{2}H(\rho),
\end{split}\ee %
where $\rho^2=r^2+R^2$ and
\be %
f(\rho)=1-\frac{1}{\rho^4},\ \ H(\rho)=1+\frac{1}{\rho^4}.
\ee %
In this coordinate, $\rho$ is the radial coordinate and horizon
is always at $\rho=1$. The Hawking temperature of black hole is
given by $T=\frac{\gamma}{\pi}$. Notice that in this background
there is also a four-form field %
\be %
 C_4=\rho^2{\rm{Vol}}_{\mathbb{R}^{3,1}}-\frac{r^2}{\rho^2}d\varphi\wedge{\rm{Vol}}_{S^3}.
\ee %

In the probe limit the dynamics of D7-branes in the
$AdS$-Schwarzschild background is described by Dirac-Born-Infeld
(DBI) action and WZ action %
\be\begin{split}\label{DBI}%
{\cal{S}}_{D7}&={\cal{S}}_{DBI}+{\cal{S}}_{WS}, \cr%
{\cal{S}}_{DBI}&=-N_f T_{D7} \int d^8 \xi \,
e^{-\phi}\sqrt{-det(\hat{g}_{ab}+(2 \pi \alpha ')
F_{ab})}, \cr %
{\cal{S}}_{WS}&=\frac{1}{2}N_f T_{D7}(2 \pi \alpha ')^2\int P[C_4]
\wedge F \wedge F,
\end{split}\ee %
where $T_{D7}^{-1}=(2\pi)^7l_s^8g_s$ is the D7-brane tension
and $\xi^a$ are worldvolume coordinates.
$\hat{g}_{ab}=g_{MN}\frac{\partial X^M}{\partial\xi^a}\frac{\partial
X^N}{\partial\xi^b}$ is the induced metric on the D7-branes and
$g_{MN}$ was introduced in \eqref{background}. $F_{ab}$ is the field
strength of the gauge fields living on the D7-branes. We use
static gauge which means that the D7-branes are extended along
$t,\vec{x},r,S^3$. $P[C_4]$ is the
pull-back of bulk four-form field to the worldvolume of D7-branes.

In order to describe the CME, we expect a current
caused by a magnetic field. We therefore consider
appropriate filed configurations on the
D7-branes as follows \cite{CME}%
\be\label{ansatz} %
 A_z(r),\ \ B=F_{xy},\ \ R(r),\ \ \varphi(t,r)=\omega t+\psi(r).
\ee %
Using the AdS/CFT dictionary, dual operator coupled to $A_z(r)$ is
$J^z$ and the expectation value of $J^z$ will describe the magnitude
of CME in the gauge theory side. $B$ is a constant external
magnetic field. Here the axial chemical potential is described by
$\omega$. In other words, the value of angular velocity of spining D7-branes
in 89-plane is identified by the axial chemical potential or more
specifically $\omega=2\mu_5$\footnote{Consider $\N=2$ SYM Lagrangian. After a chiral rotation $\psi\rightarrow e^{-i\gamma^5\varphi/2}\psi$,
the following new term appears in the fermion's kinetic term
\be %
 -\frac{\partial_\mu\varphi}{2}\bar{\psi}\gamma^\mu\gamma^5\psi.\nonumber
\ee %
 Using $\varphi=\omega t$, it is evidently seen that $\omega=2\mu_5$(for more details see \cite{CME}).}.

Inserting the above ansatz in \eqref{DBI}, the density action for the D7-branes becomes %
\be\begin{split}\label{DBIaction}  %
S_{DBI}&= -{\cal{N}} \int dr  \, g_{ss}^{3/2}g_{xx}^{3/2}\sqrt{1 +
\frac{(2\pi\alpha')^2B^2}{g_{xx}^2}} \, \cr %
&\times \sqrt{(|g_{tt}|-g_{\psi\psi} \omega^2)(g_{rr}+g_{RR}\,R'^2
+ g^{xx} A_z'^2)+|g_{tt}|g_{\psi\psi}\psi'^2}, \cr %
S_{WZ} &= -{\cal{N}} (2\pi\alpha')B \omega \int dr \, \frac{r^4}{\rho^4} A_z',
\end{split}\ee %
where $'=\frac{\partial}{\partial r}$ and ${\cal{N}}=\frac{\lambda
N_f N_c}{(2\pi)^4}$. Note that for all the probe branes the same electrical charge has been supposed.
Taking the integral over the worldvolume of
D7-branes we get the infinite volume of Minkowski space,
$V_{\mathbb{R}^{3,1}}$, and the volume of $S^3$. We absorb the volume of
$S^3$ in ${\cal{N}}$ and define a density action as
$S_{D7}=\frac{{\cal{S}}_{D7}}{V_{\mathbb{R}^{3,1}}}$. Notice that
the WZ action is not zero and in fact it plays the role of the anomaly
in the gauge theory side.

As it is clear from \eqref{DBIaction}, the action depends only on
the derivative of $A_z$ and $\psi$ and we therefore have two constants
of motion
\textit{i.e.} %
\be\begin{split}\label{constants} %
 \frac{\delta S_{D7}}{\delta\psi'}&=\alpha, \cr
 \frac{\delta \hat{S}_{D7}}{\delta A'_z}&=\beta,
\end{split}\ee %
where the hat means that a Legendre-transformation has been applied. Applying two Legendre-transformations
with respect to $\psi'$ and $A'_z$,
we have (see Appendix A)%
\be\begin{split}\label{actionAfterLeg} %
\hat{\hat{S}}_{D7}&=-{\cal{N}}\int dr r^2\sqrt{1+ R'^2}
\sqrt{a(r)b(r)-c(r)^2}, %
\end{split}\ee %
where %
\be\begin{split} %
 a(r)&=|g_{tt}|-g_{\psi\psi}\omega^2, \cr %
 b(r)&=1 + \frac{(2\pi\alpha')^2B^2}{g_{xx}^2}-\frac{\a^2/\N^2}{|g_{tt}|g_{\psi\psi} g_{xx}^3g_{ss}^3}, \cr %
 c(r)&=\frac{\beta}{\N} + (2\pi\alpha')B \omega g_{ss}^2.
\end{split}\ee %
Using the above action the equation of motion for
$R(r)$ can be derived. This equation is complicated to solve analytically and
hence we use numerical method.

According to the AdS/CFT correspondence, for different fields
leading and sub-leading terms in near boundary asymptotic expansion
define a source for the dual operator and its expectation value,
respectively. For $R(r)$ the leading term corresponds
to the mass of the ${\cal{N}}=2$ fundamental matter and the
sub-leading term is $\langle O_m\rangle$ where $O_m$ is the dual
operator to mass, \textit{i.e.}
\be %
 R(r)=|m|+\frac{\langle O_m\rangle}{r^2}.
\ee %
In \cite{CME} it was obtained that the expectation values of dual
operators to $A_z$ and $\varphi$ are \footnote{Note that the leading
term in $\varphi$ is $\omega t$. Moreover in the first reference of
\cite{Rebhan}, it was discussed that the leading term in $A_z$
corresponds to a meson gradient. In our case, as it is clear from
\eqref{current}, the value of CME is independent of the leading term
in $A_z$.}
\bse\begin{align} %
 \label{current}\langle J^z\rangle &=-(2\pi\alpha')\beta,\\
 \label{CT} \langle O_\varphi\rangle &=\alpha.
\end{align}\ese %
Therefore $\beta$, up to a constant, gives the value of CME in the
gauge theory side. Also regarding the discussion about discrete
spacetime symmetries, $\alpha$ is an order parameter of spontaneous
CT symmetry breaking \cite{CME}.

By requiring that the action \eqref{actionAfterLeg} is real,
two consistent options can be found \cite{CME}. The first one is
when $a(r)$ does not change sign between $r\rightarrow 0$ and
$r\rightarrow\infty$. Since to $a(r)>0$ at infinity, reality
condition of the action imposes $\alpha=0$ and $\beta=0$. From
\eqref{current} it is clearly seen that in this case there is no
CME. The second option is when $a(r)$ does change sign at $r=r_*$
\textit{i.e.} %
\be\label{condition1} %
 a(r_*)=0,
\ee %
and again the reality condition imposes
\be\label{condition2} %
 b(r_*)=0,\ \  c(r_*)=0.
\ee %
Hence \eqref{condition1} and \eqref{condition2} respectively lead to %
\bse\label{ahmad2}\begin{align} %
 \label{Ahmad1}f(\rho_*)^2&=\frac{2H(\rho_*)}{\gamma^2}\frac{\omega^2R_*^2}{(r_*^2+R_*^2)^2},\\
 \label{alpha}\alpha&=-\frac{{\cal{N}}\gamma^4}{4}r_*^3R_*f(\rho_*)H(\rho_*)\sqrt{1+\frac{4(2\pi\alpha')^2B^2}{\gamma^4(r_*^2+R_*^2)^2H(\rho_*)^2}},\\
 \label{beta}\beta&=-(2\pi\alpha'){\cal{N}}B\omega\frac{r_*^4}{(r_*^2+R_*^2)^2},
\end{align}\ese %
where $R_*=R(r_*)$ and $\rho_*^2=r_*^2+R_*^2$. The equation \eqref{Ahmad1} describes the worldvolume horizon on the
D7-branes. We now compare various properties of different choices for $\alpha$ and $\beta$ \cite{CME}.
\begin{enumerate}
  \item $\alpha\neq0,\ \beta\neq0$ \\
  According to \eqref{current} and \eqref{CT}, the CME exists and CT symmetry is broken.
  In this case the mass dependence of CME can be studied and as
  the red curve in Fig. \ref{B-m}(right) shows the value of CME goes to zero for sufficiently large $m$.
  \item $\alpha=0,\ \beta\neq0$ \\
  This solution describes a massless solution defined by $R(r)=0$ and $\varphi=\omega
  t$. From \eqref{current} and \eqref{beta}, the value of CME is %
  \be\label{CME} %
  \langle J^z\rangle=(2\pi\alpha')^2{\cal{N}}B\omega=\frac{N_fN_c}{2\pi^2}\mu_5B
  \ee %
  Here CT symmetry has been restored.
  \item $\alpha\neq0,\ \beta=0$ \\
  Imposing $\beta=0$ leads to $\alpha=0$ and therefore there is no such a family of solutions.
  \color{black}
  \item $\alpha=0,\ \beta=0$ \\
  In this case the value of CME is zero and CT symmetry is restored.
\end{enumerate}

Before closing this section we would like to explain the numerical
method for solving the equation of motion $R(r)$. For doing that, we
need two boundary conditions at a specific point. When $a(r)>0$, as
it was already mentioned, we have $\alpha=\beta=0$. In this case we
use the following boundary conditions. The first boundary condition
is $R(0)=R_0$ which $R_0$ must be always greater than $\rho_*(r=0)$,
otherwise $a(r)$ vanishes at $r=r_*$. The second one is $R'(0)=0$
which guarantees the regularity of D7-branes. In the case of
$a(r_*)=0$, by choosing $R_*$ and $r_*$ satisfying in
\eqref{Ahmad1}, the values of $\alpha$ and $\beta$ can be found by
using \eqref{alpha} and \eqref{beta}. Then the first boundary
condition is $R_*=R(r_*)$. Although the second boundary condition,
$R'(r_*)$, can be obtained by using the equation of motion, it is
numerically seen that the equation of motion for $R(r)$ is
insensitive to $R'(r_*)$.

\section{$\alpha'$-Corrected CME}
Since the AdS/CFT correspondence refers to the complete string theory one
can consider stringy correction to the ten dimensional
supergravity action. It is well-known that in extremal case the metric does not change
\cite{correction} and hence the value of CME is still given by
\eqref{CME}. But for non-extremal case the first order in the
weakly curved type IIb background occurs at $O(\alpha'^3)$
\cite{correction1} which is dual to $O(\lambda^{-\frac{3}{2}})$ in
the gauge theory side.
In this section we will study the effect of $\alpha'$-correction
on $\alpha$, $\beta$, the worldvolume horizon on D7-branes and  $R(r)$ corresponding to the different
D7-brane embeddings in the ten dimensional $\alpha'$-corrected background.

The $\alpha'$-corrected metric is \cite{Gubser}
\be\begin{split} %
 ds^2&=G_{tt}dt^2+G_{xx}d\vec{x}^2+G_{uu}du^2+G_{ss}d^2\Omega_5, \cr
 d\Omega_5^2&=d\theta^2+\sin^2\theta d\varphi^2 + \cos^2 \theta
 d\Omega_3^2,
\end{split}\ee %
where the metric components are given by ($w=\frac{u_0}{u}$), %
\be\begin{split} %
 -G_{tt}&=u^2(1-w^{4})T(w),\cr
 G_{xx}&=u^2 X(w),\cr
 G_{uu}&=u^{-2}(1-w^{4})^{-1}U(w),\cr
 G_{ss}&=1+2S(w),
\end{split}\ee %
and %
\be\begin{split} %
 T(w)&=1-k(75w^{4}+\frac{1225}{16}w^{8}-\frac{695}{16}w^{12})+O(k^2), \cr
 X(w)&=1-\frac{25k}{16}w^{8}(1+w^{4})+O(k^2), \cr
 U(w)&=1+k(75w^{4}+\frac{1175}{16}w^{8}-\frac{4585}{16}w^{12})+O(k^2), \cr
 S(w)&=\frac{15k}{32}w^{8}(1+w^{4})+O(k^2). %
\end{split}\ee %
The expansion parameter $k$ in terms of the t'Hooft coupling constant is
\be %
 k=\frac{\zeta(3)}{8} \lambda^{-3/2},
\ee %
where $\zeta$ is the Riemann zeta function. The correction in the
t'Hooft coupling constant corresponds to the $\a'$-correction on the
gravity side. The corrected metric has an event horizon at $u=u_0$.
At large $u$ the above geometry becomes $AdS_5\times S^5$. The
Hawking temperature is now
\be %
 T=\frac{u_0}{\pi(1-k)}.
\ee %
In the presence of $\alpha'$-correction, the Dilaton field is not a constant
anymore and it depends on $u$ as
\be %
 \phi(w)=-\log
 g_s-\frac{45k}{8}(w^{4}+\frac{1}{2}w^{8}+\frac{1}{3}w^{12})+O(k^2).
\ee %

In order to get a suitable form of the metric we apply two
successive changes of coordinate as
\be %
 u^2=\rho^2(1+\frac{\rho_0^4}{\rho^4}),
\ee %
and
\be\begin{split}\label{changeofcoordinate} %
 r&=\rho \sin \theta, \cr
 R&=\rho \cos \theta.
\end{split}\ee %
The form of $\alpha'$-corrected metric is finally given by %
\be %
 ds^2=-|g_{tt}|dt^2+g_{xx}d\vec{x}^2+g_{rr}dr^2+g_{RR}dR^2+g_{rR}drdR+g_{ss}ds_{S^3}^2+g_{\varphi\varphi}d{\varphi}^2
\ee %
where its components up to second order of $k$ are\footnote{In this coordinate $\gamma=u_0$.}
\be\begin{split}\label{eq001} %
|g_{tt}|&=\rho ^2 \frac{\gamma ^2 f^2(\rho)}{2 H(\rho)}-\frac{5
\gamma ^2 f^2(\rho) U_1(\rho) }{2 \rho ^ {18} H^7(\rho)} k,
\hspace{.3cm} g_{rR}= \frac{ 40 r R}{\rho ^{24}} \frac{U_3(\rho)}{H^6(\rho)} k,  \cr %
g_{xx}&= \rho ^2\frac{\gamma^2}{2}  H(\rho)-\frac{25 \gamma ^2
T(\rho)}{2 \rho ^{14} H^5(\rho)}k,\hspace{1.0cm} g_{ss}=
\frac{r^2}{\rho ^2}+\frac{15 r^2 T(\rho) }{\rho ^{18}H^6(\rho)}k,  \cr %
g_{rr}&=\frac{1}{\rho ^2}+\frac{15  R^2\rho ^4  T(\rho)+ 5 r^2
U_2(\rho)}{\rho ^{24}H^6(\rho)} k,\hspace{.3cm} g_{\varphi\varphi}=
\frac{R^2}{\rho ^2} +\frac{15 R^2 T(\rho)}{\rho^{18}H^6(\rho)}k, \cr %
g_{RR}&= \frac{1}{\rho ^2}+\frac{15  R^2\rho ^4  T(\rho)+ 5 r^2
U_2(\rho)}{\rho ^{24}H^6(\rho)} k.
\end{split}\ee %
Note that $\rho^2=r^2+R^2$,
\be\begin{split}\label{eq002}%
U_1(\rho)&=60+485 \rho ^4+294 \rho ^8+485 \rho ^{12}+60 \rho^{16},\cr %
U_2(\rho)&=60+475 \rho^4-2838 \rho ^8+475 \rho ^{12}+60 \rho ^{16}, \cr %
U_3(\rho)&=15+118 \rho ^4-714 \rho ^8+118 \rho ^{12}+15\rho^{16},\cr %
T(\rho)&= 1+6 \rho ^4+\rho ^8,
\end{split}\ee %

\begin{figure}[ht]
\begin{center}
\includegraphics[width=2.6 in]{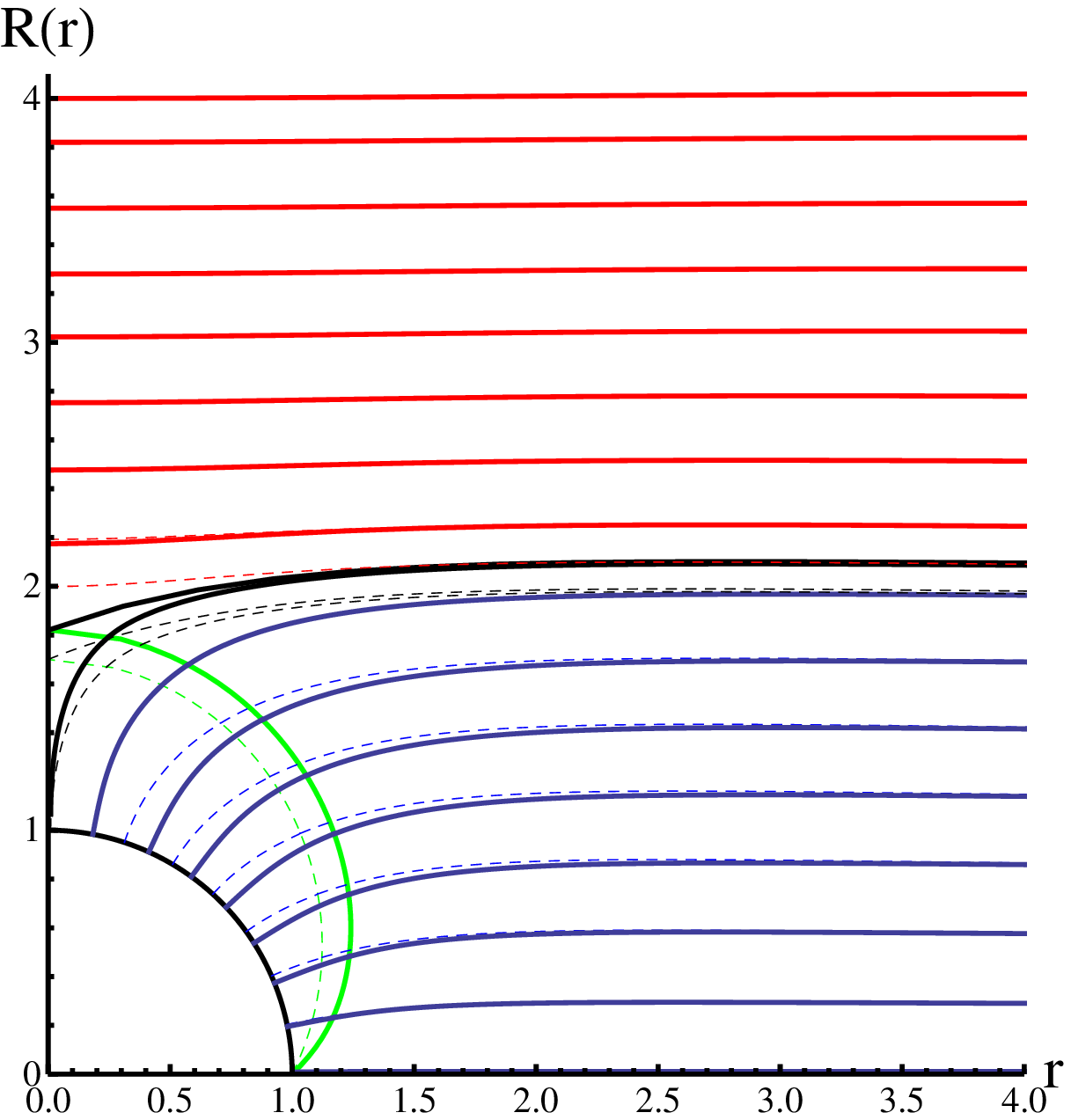}
\hspace{2mm}
\includegraphics[width=2.6 in]{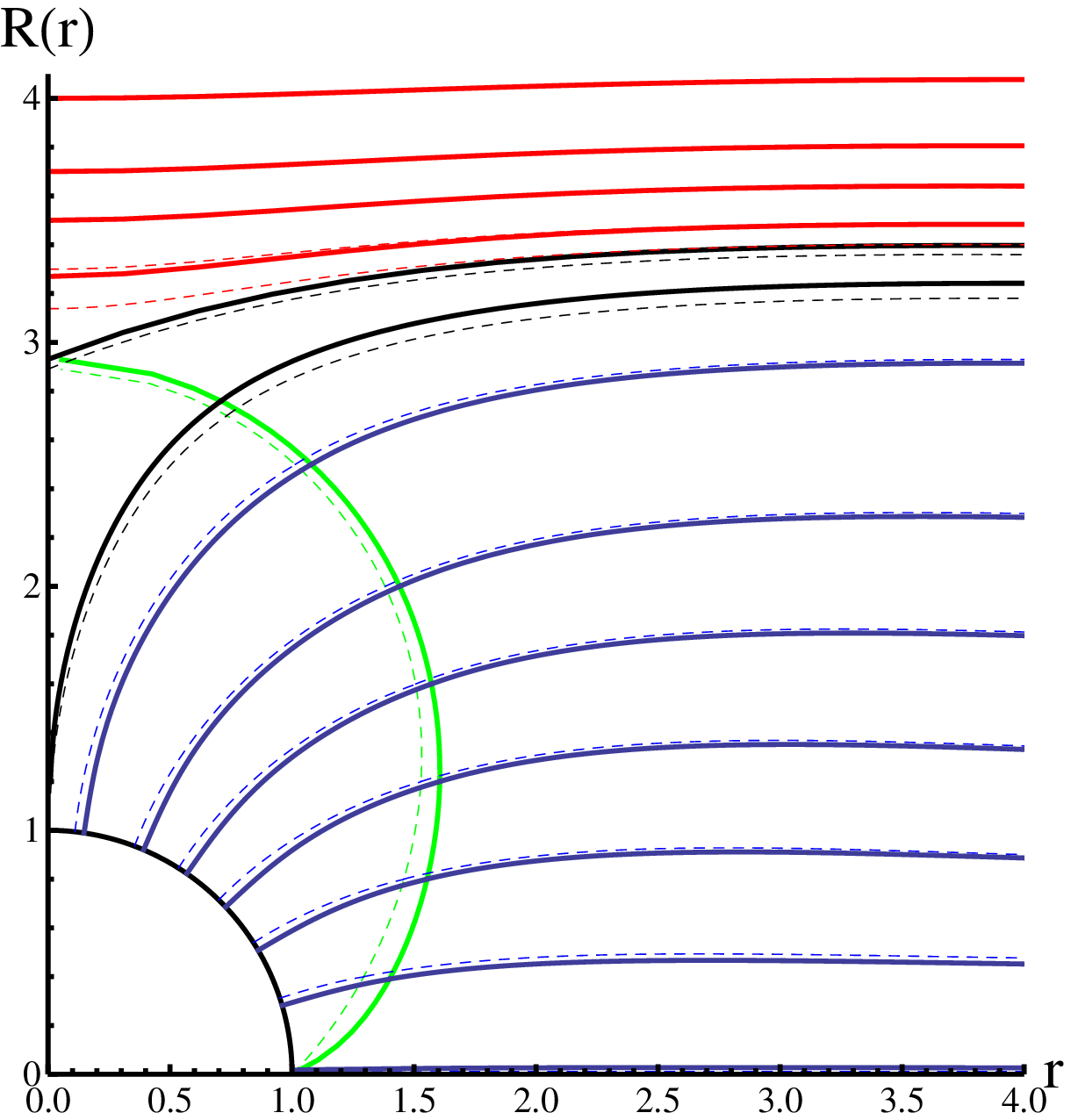}
\caption{Numerical D7-brane embeddings for $k=0$ (dashed curves) and
$k=0.007$ (continuous curves) corresponding to $T=0.5\pi^{-1}$(right)
and $T=\pi^{-1}$(left). Green curves show the worldvolume horizon
($B=1$ and $\omega=1$).\label{alphaBeta}}
\end{center}
\end{figure}%

and also $3\rho^4T(\rho)+4U_3(\rho)=U_2(\rho)$.
Now the density action for the D7-branes is %
\be\begin{split}\label{DBIaction2} %
S_{DBI}&= -{\cal{N}} \int dr  \, e^{-\phi}\,
g_{ss}^{3/2}\sqrt{1 + \frac{(2\pi\alpha')^2B^2}{g_{xx}^2}}
g_{xx}^{3/2} \, \cr %
&\times \sqrt{(|g_{tt}|-g_{\psi\psi} \omega^2)(g_{rr}+g_{RR}\,R'^2
+g_{rR}R'+ g^{xx} A_z'^2)+|g_{tt}|g_{\psi\psi}\psi'^2}, \cr %
S_{WZ} &= -{\cal{N}} (2\pi\alpha')B \omega \int dr \, \frac{r^4}{\rho^4} A_z',
\end{split}\ee %
where \eqref{ansatz} has been used. Notice that the four-form field
in the background and consequently $S_{WS}$ are not corrected
\cite{Gubser}. Comparing with \eqref{DBIaction}, the
$\alpha'$-corrected action has two extra terms. The first term is
$e^{-\phi}$ which is not a constant anymore and the second one is
$g_{rR}R'$ . As before we have two constants of motion,
\textit{i.e.} $\alpha$ and $\beta$. $\psi'$ and $A'_z$ can be
eliminated in favor of $\alpha$ and $\beta$ by two successive
Legendre-transformations and the action finally becomes \eqref{seq
Legen}. Additional details can be found in Appendix A. Similar to
the previous section we numerically solve the equation of motion for
$R(r)$ and the resulting solutions are shown in Fig.
\ref{alphaBeta}.

\begin{figure}[ht]
\begin{center}
\includegraphics[width=3 in]{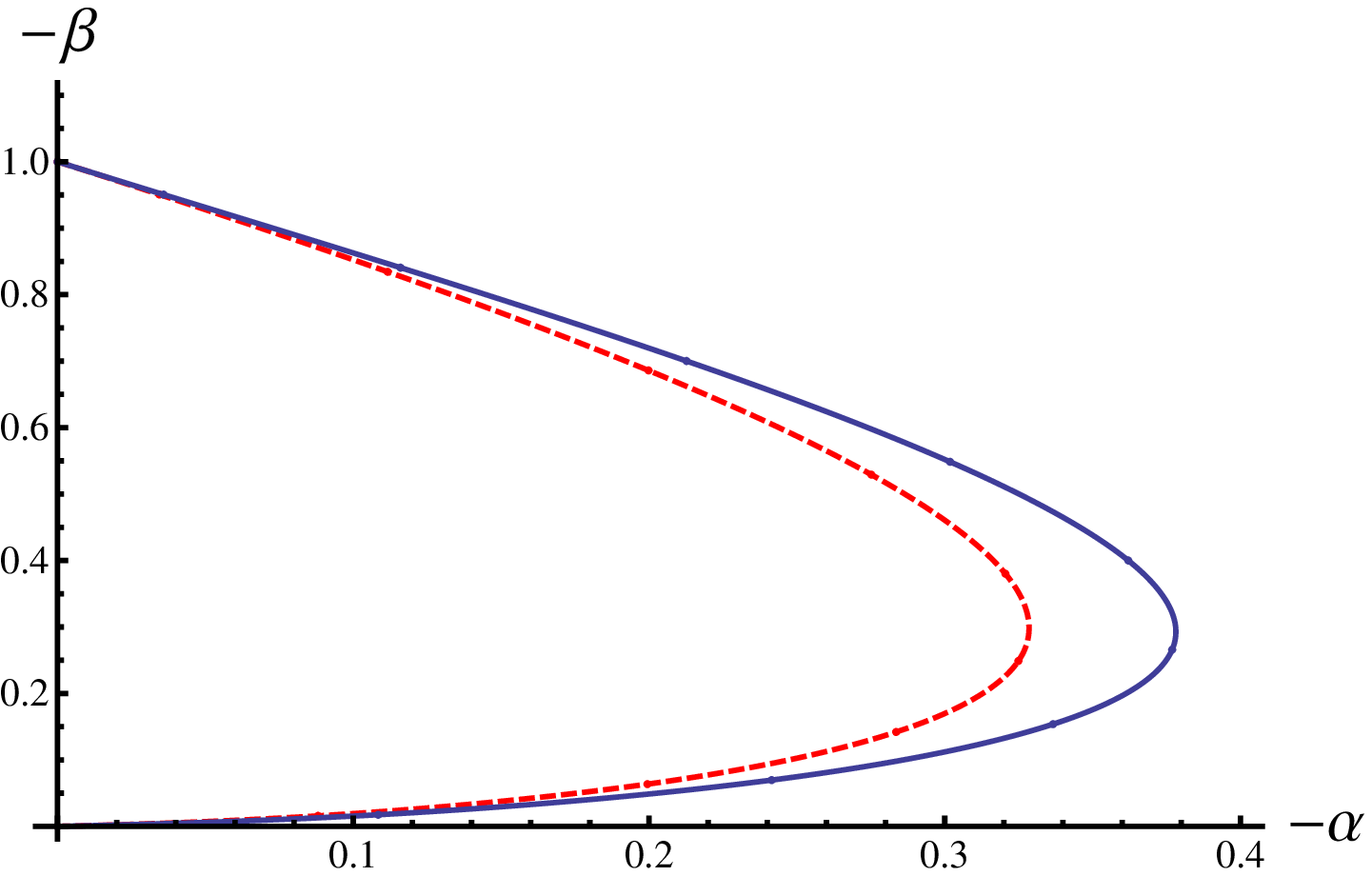}
\hspace{2mm}
\includegraphics[width=1.7 in]{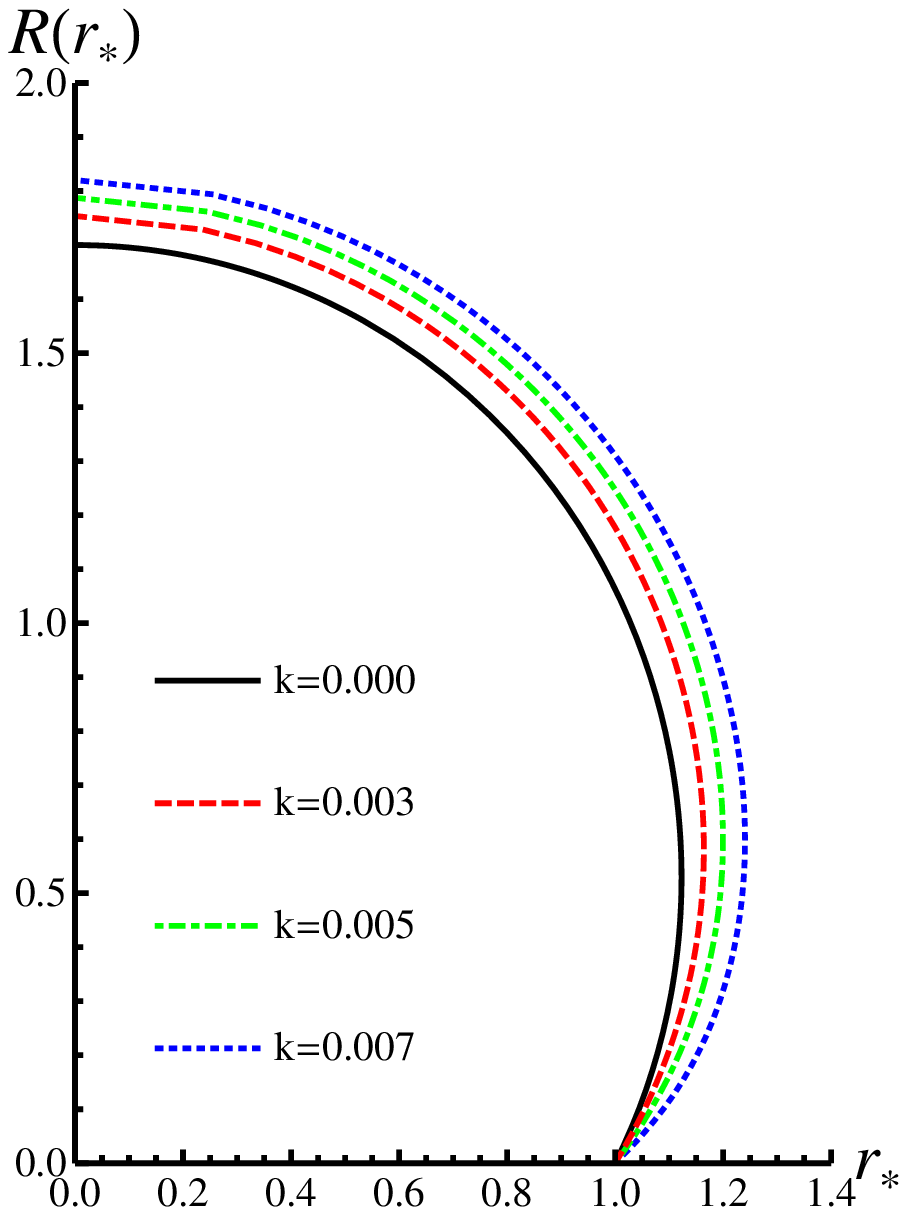}
\caption{(Left) The blue, continuous curve shows $\alpha'$-corrected
case ($k=0.007$). (Right) The worldvolume horizon on the D7-branes
for different values of $k$ ($B=1$, $\omega=1$ and $T=\pi^{-1}$)
.\label{worldvolumehorizon}}
\end{center}
\end{figure}

The position of $\alpha'$-corrected worldvolume horizon on the
D7-branes, $\rho_*$, can be found by solving
\eqref{worldvolumehorizonk}. In Fig. \ref{worldvolumehorizon}(right)
we assume different values for $k$ and plot the worldvolume horizon
in terms of $k$. It is obviously seen that by increasing $k$ the
value of $\rho_*$ also increases. Moreover notice that $R(r_*=1)$ is
zero for both worldvolume and AdS-Schwarzschild horizon meaning that
these two horizons are coincidence at $r_*=1$ for  small values of
$k$. This fact can be obtained by using \eqref{worldvolumehorizonk}.
Setting $R_*=0$ we have %
\be\label{farid} %
\frac{\gamma^2}{2}\frac{f(r_*)^2}{H(r_*)}\left(1 -\frac{5k
U_1(r_*)}{r_*^{20}H(r_*)^6}\right)=0.
\ee %
For permitted $k$, the term in the parentheses does not vanish and
therefore $f(r_*)$ must be zero leading to $r_*=1$.

Each point on the worldvolume horizon, to be more specific $r_*$ and
$R_*$, fixes a specific value for $\alpha$ and $\beta$. The dashed,red
curve in Fig. \ref{worldvolumehorizon}(left) shows that for a
given value of $\alpha$ one gets two $\beta$s. In the simplest case
( \textit{i.e.} assuming $\alpha=0$) we find two options: $r_*=0$ or
$R_*=0$ (see \eqref{alpha} and \eqref{beta}). The former describes a
solution with $\beta=0$ and the latter is a massless solution.
Therefore the minimum and the maximum values of CME are given by
$\alpha=0$. Since $\alpha$ ia equal to zero these solutions are CT
invariant. Notice that in this figure the value of $\beta$ is
normalized to the value in \eqref{CME}. \eqref{ahmad3} and
\eqref{ahmad4} show that the same argument is also true for the
$\alpha'$-corrected background (the continuous, blue curve in the
Fig. \ref{worldvolumehorizon}(left)).

\begin{figure}[ht]
\begin{center}
\includegraphics[width=2.6 in]{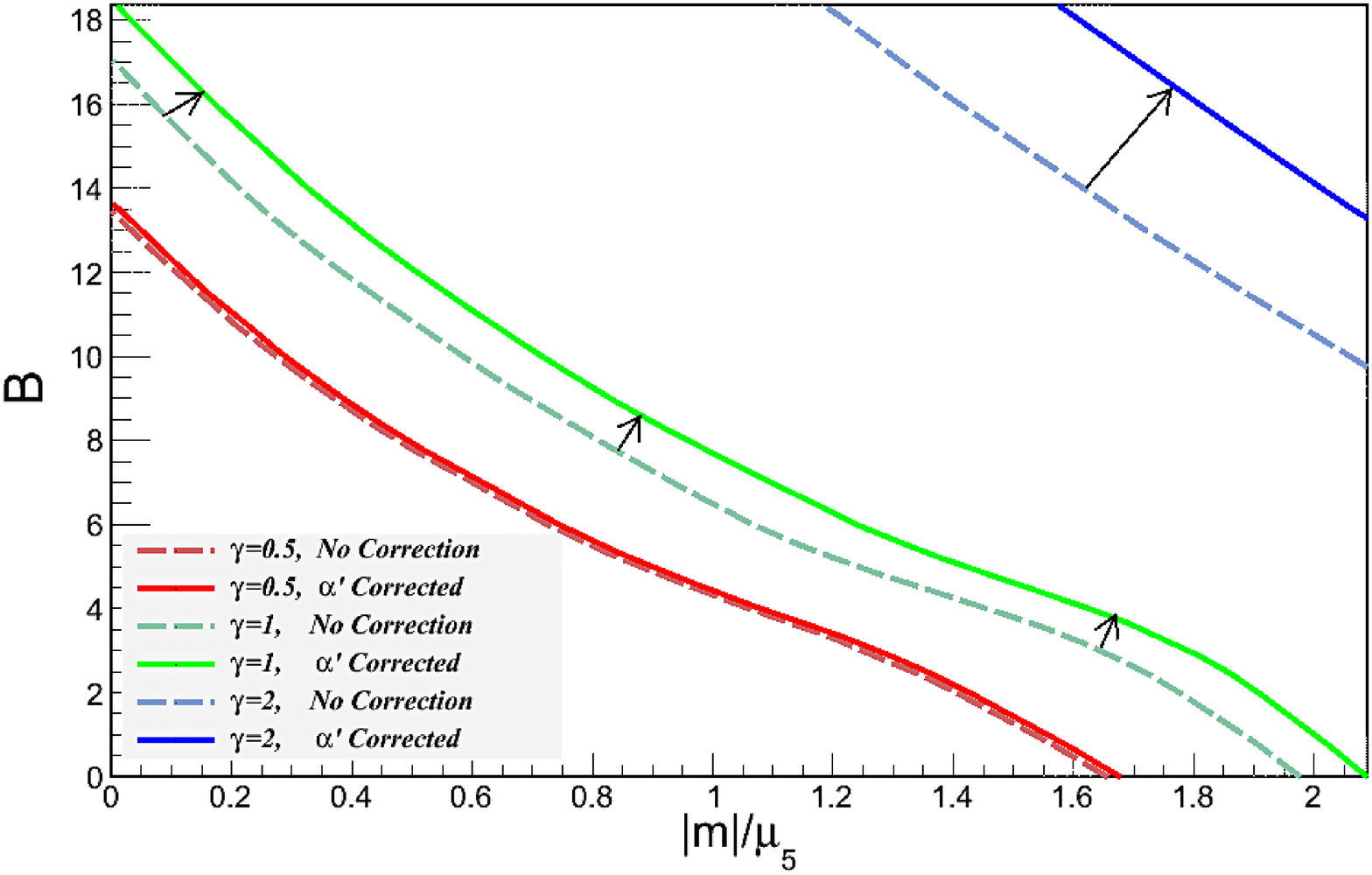}
\hspace{2mm}
\includegraphics[width=2.6 in]{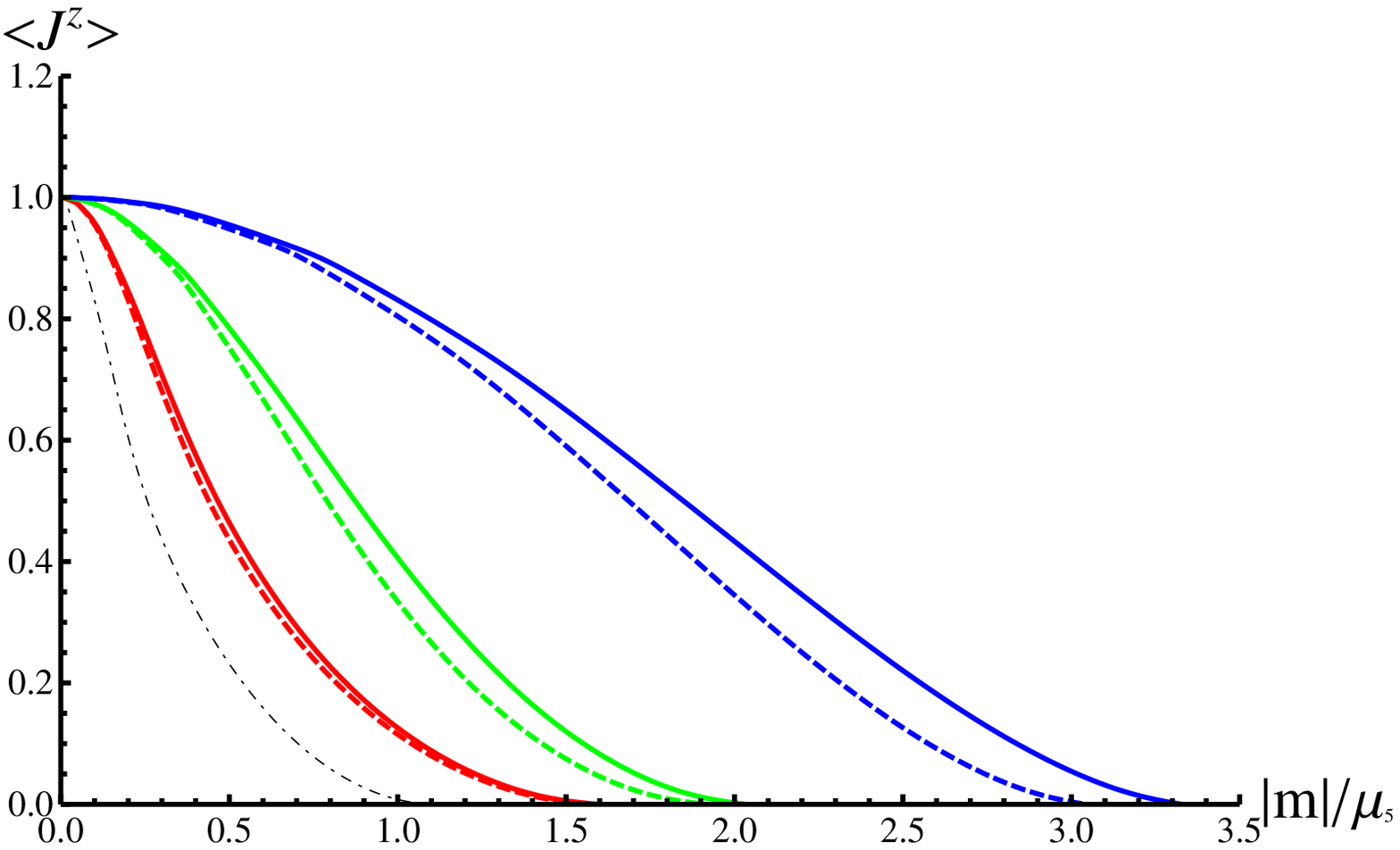}
\caption{ The red, green and blue continuous curves show the
$\alpha'$-corrected cases for $T=0.5\pi^{-1},\pi^{-1}$ and $2\pi^{-1}$, respectively
($\omega=1$). (Left) The left part of each lines in $B-m/\mu_5$ plot
shows the region in which $\langle J^z\rangle$ is not zero. (Right)
The value of $\langle J^z\rangle$ as a function of $m/\mu_5$.  The
black curve shows the zero temperature case. \label{B-m}}
\end{center}
\end{figure}

Using the action \eqref{seq Legen} one can compute the values of
$\alpha$ and $\beta$ and also the position of worldvolume horizon.
The resulting expressions for $\alpha$ and the worldvolume horizon are so
lengthy (see Appendix A) and hence we here explicitly state the
value of $\beta$
\be\label{NCME} %
 \beta =-(2\pi\alpha')\N B \omega \frac{ r_*^4 }{\rho_*^4}.
\ee %
The value of $\beta$ does not explicitly depend on
$\alpha'$-correction. The reason is that the WZ term in the action
is not affected by $\alpha'$-correction. Since the position of
worldvolume horizon varies, the magnitude of $\beta$ also changes.
As it is clear from Fig. \ref{B-m}(right), the value of CME, $\langle
J^z\rangle=-(2\pi\alpha')\beta$, rises by $\alpha'$-correction for
the massive solutions and this increase is more considerable for
higher temperatures. Note also that as it was discussed in
\eqref{farid} this value dose not vary for the massless solutions.

\begin{figure}[ht]
\begin{center}
\includegraphics[width=2.5 in]{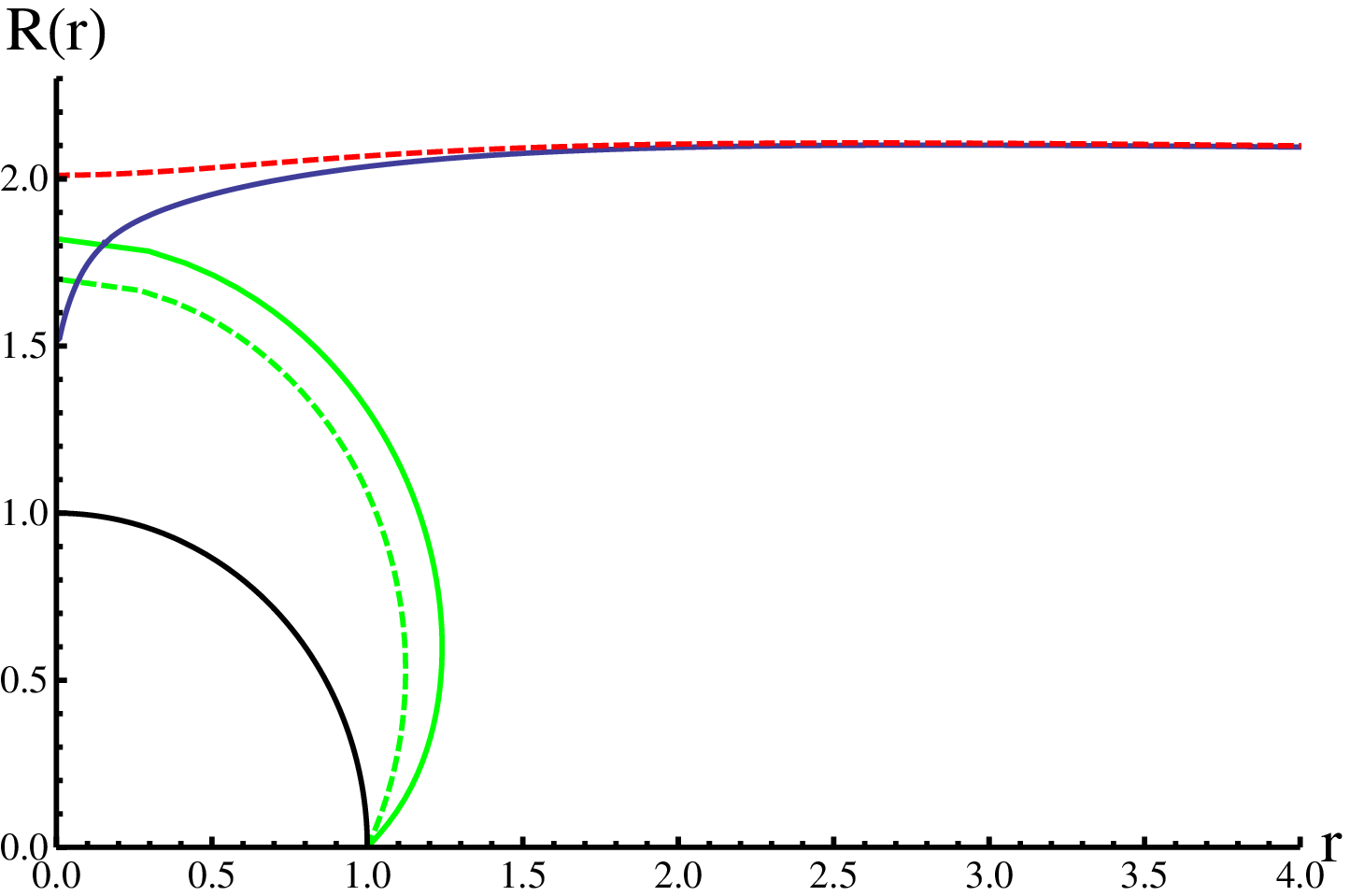}
\hspace{3mm}
\includegraphics[width=2.5 in]{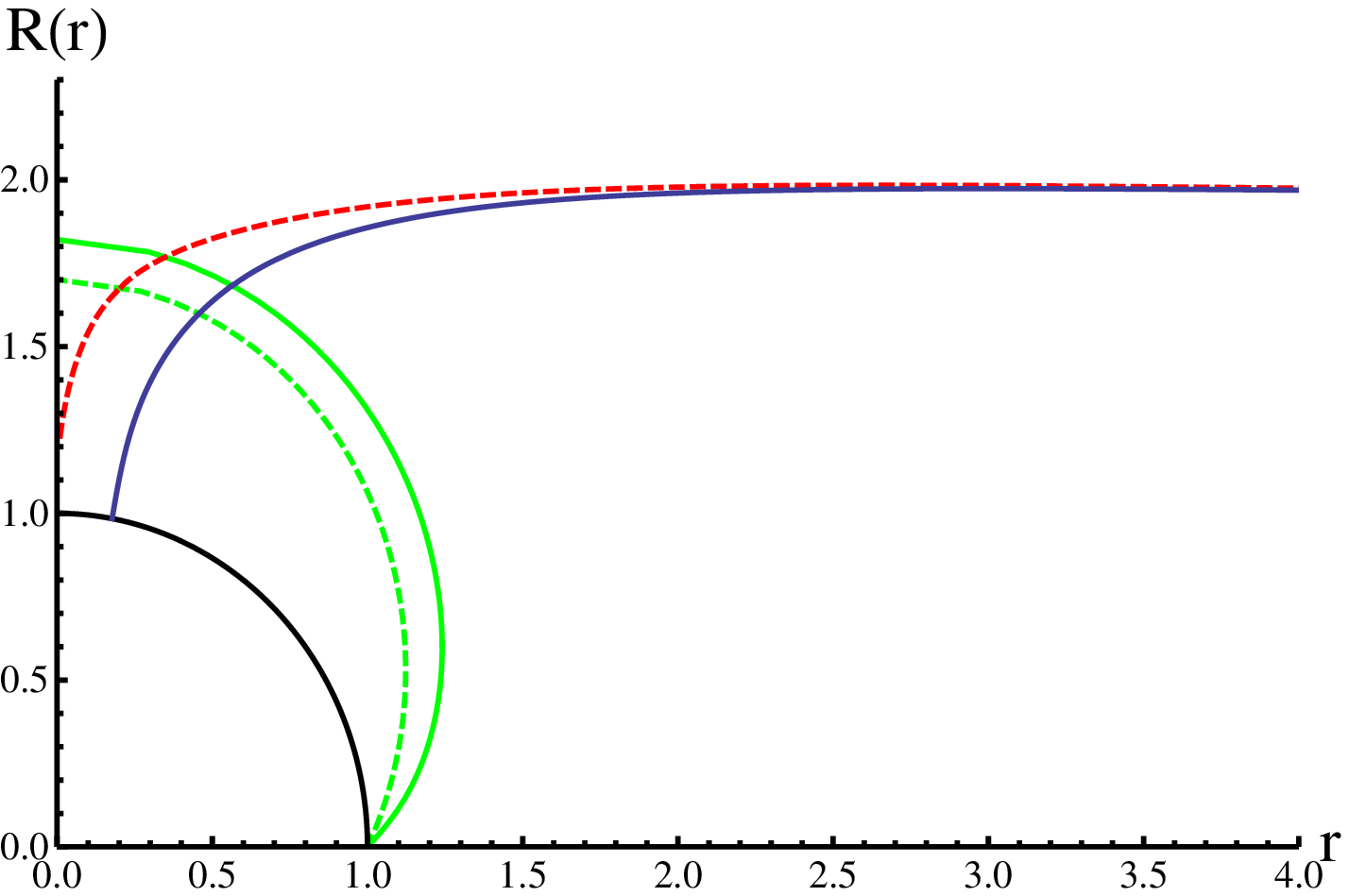}
\caption{(Left) In the case of $m=2.037$, a Minkowski embedding
without horizon for $k=0$ modifies to a Minkowski embedding with
horizon for $k=0.007$. This happens for a range of masses between
$m=2.028$ and $m=2.038$. (Right) A Minkowski embedding with horizon
for $k=0$ modifies to a black hole embedding for $k=0.007$ when the
value of $m$ is fixed by $1.912$. It happens for a range of masses
between $m=1.907$ and $m=1.919$ ($B=1$, $\omega=1$ and
$T=\pi^{-1}$).\label{BH}}
\end{center}
\end{figure}

At finite temperature the various embeddings of D7-branes can be
classified into three groups. The Minkowski embeddings are those
embeddings where the probe D7- branes close off above the horizon.
In other words, the $S^3$ part of D7-branes shrinks to zero at an
arbitrary value $\rho_s=R(0)>1$. Also $\rho_s$ can be bigger than
the value of worldvolume horizon on the D7-brane, \textit{i.e.}
$\rho_s>\rho_*$. This group of embeddings is called
\textit{Minkowski embeddings without horizon}
\cite{Kobayashi,Mateos}. In this case $\alpha$ and $\beta$ are equal
to zero and there is no CME. Conversely if $\rho_s<\rho_*$, the
solutions are called \textit{Minkowski embeddings with horizon}. The
third group is \textit{black hole embeddings}
\cite{Kobayashi,Mateos}. In this group the $S^3$ part of D7-branes
shrinks but does not reach zero size for $\rho_s\geq1$. These three
groups are shown in Fig. \ref{alphaBeta} by red, black and blue
curves respectively. Green curves show the worldvolume horizon in
this figure. The continuous curves show the $\alpha'$-corrected
embeddings of D7-branes. As is clear from Fig. \ref{alphaBeta}, for
the Minkowski embeddings with horizon $R'(0)$ is not zero and
therefore such D7-branes have conical singularity (for more detail
see \cite{CME}).

In Fig. \ref{BH}(left) the red, dashed curve shows a Minkowski
embedding without horizon for a specified value of $m$ when the
correction is not considered and the value of CME is clearly zero
for this embedding. An interesting result is that in the presence of
$\alpha'$-correction the embedding becomes a Minkowski embedding
with horizon (the blue, continuous curve) for the same value of mass
and therefore we have a non-zero value for the CME. In other words,
although before applying $\alpha'$-correction $a(r)$ was always
bigger than zero for the all values of $r$, after including
$\alpha'$-correction it becomes zero at $r=r_*$ and then a worldvolume horizon appears.
Similarly Fig. \ref{BH}(right) shows that a Minkowski embedding with
horizon (the dashed, red curve) is modified into a black hole embedding
(the blue, continuous curve) and as a result the conical singularity
is removed by the effect of $\alpha'$-correction. Shortly, the
number of D7-brane embeddings, which gives a non-zero value for the
CME, increases in the presence of $\alpha'$-correction.

The massless solution exists for all values of the magnetic field.
However, for massive solutions Fig. \ref{B-m}(left) shows that for a
given value of the magnetic field there is a maximum value for
$\frac{m}{\mu_5}$. In the region with
$\frac{m}{\mu_5}<(\frac{m}{\mu_5})_{max}$ the value of CME is
non-zero which is described by black hole embeddings or Minkowski
embeddings with horizon . For
$\frac{m}{\mu_5}>(\frac{m}{\mu_5})_{max}$ Minkowski embeddings
without horizon exist and hence $\langle J^z\rangle=0$. As it is
clear from Fig. \ref{B-m}(left) the $\alpha'$-correction increases
the value of $(\frac{m}{\mu_5})_{max}$. For higher temperatures the
value of $(\frac{m}{\mu_5})_{max}$ rises. Similarly there is a
$B_{max}$ for a given value of $\frac{m}{\mu_5}$ and for
$B<B_{max}$, $\langle J^z\rangle\neq0$. The $\alpha'$-correction
also increases the value of $B_{max}$.

The value of \eqref{NCME} can be numerically evaluated. The
corresponding plot is presented in Fig. \ref{B-m}(right). In the
massless limit the magnitude of CME, normalized to the value in
\eqref{CME}, has a maximum value and then by increasing
$\frac{m}{\mu_5}$ the value of CME decreases and reaches zero at
$(\frac{m}{\mu_5})_{max}$. The continuous curves correspond to the
$\alpha'$-corrected cases. As it is clear from this figure the
effect of $\alpha'$-correction is more sensible at higher
temperatures. Note also that in the case of zero temperature since
the $\alpha'$-correction does not change the $AdS_5\times S^5$
background, the value of CME is still given by \eqref{CME}.

\section{Conclusion}
Our results can be summarized as follows.
\begin{itemize}
  \item The main result is that for massive solutions the value of CME always rises in the presence of
        $\alpha'$-correction. For higher temperatures this increase is more significant.
        In the case of massless solution this value does not change.
  \item A family of D7-brane embeddings, Minkowski emmbeddings without
        horizon, describes solutions with no CME. The $\alpha'$-correction
        changes these embeddings to the Minkowski emmbeddings with horizon
        where the value of CME is not zero anymore. In other words, the
        embeddings can be changed by the effect of $\alpha'$-correction. It
        is important to notice that this happens for a range of $(\frac{m}{\mu_5})$s.
  \item Similarly, the Minkowki embeddings with horizon change into the
        black hole embeddings for a range of $(\frac{m}{\mu_5})$s in presence of the
        $\alpha'$-correction. In this case both embeddings have a non-zero
        value for the CME. But the singularity of Minkowski embeddings with
        horizon is removed by the effect of $\alpha'$-correction.
\end{itemize}

\section*{Acknowledgements}
We thank very much F. Ardalan, H. Arfaei, K. Bitaghsir, A. Davody,
A. E. Mosaffa, A. O'Bannon and M. M. Sheikh-Jabbari for discussions.
M. A. would also like to thank H. Ebrahim for many useful
discussions.

\section*{Appendix A}

In order to find the equation of motion for $R(r)$, we apply two successive Legendre-transformations. The first Legendre-transformation is \be\label{30}
\hat{S}_{D7}=S_{D7}-\int dr \psi' \frac{\delta S_{D7}}{\delta \psi'}.
\ee
where $S_{D7}$ is given by \eqref{DBIaction2}. $S_{WZ}$ does not depend on the $\psi$ and hence \eqref{30} can be written as
\bea\label{Ahmad}
\hat{S}_{DBI} & = & S_{DBI} - \int dr \psi' \, \frac{\delta S_{DBI}}{\delta \psi'}  \\
& = & -\N \int dr \, g_{ss}^{3/2} g_{xx}^{3/2}\sqrt{|g_{tt}|-g_{\psi\psi} \omega^2}  \nonumber\\
& \times & \sqrt{g_{rr}+g_{RR}\,R'^2 +g_{rR}R'+ g^{xx} A_z'^2}\sqrt{1 +
\frac{(2\pi\alpha')^2B^2}{g_{xx}^2} - \frac{\a^2/\N^2}{|g_{tt}| g_{\psi\psi}
g_{xx}^3 g_{ss}^3}}, \nonumber \eea
where $\alpha$ is introduced in \eqref{constants}. The second Legendre-transformation with respect to $A'_z$ is
\bea
\label{seq Legen}
\hat{\hat{S}}_{D7} & = & \hat{S}_{D7} - \int dr A_z' \frac{\delta \hat{S}_{D7}}{\delta A_z'} \\
& = & -\N \int dr \sqrt{g_{rr} + g_{RR} R'^2+g_{rR}R'} \nonumber \\
&\times & \Bigg[g_{xx}^3 g_{ss}^3 \left(|g_{tt}|-g_{\psi\psi}
\omega^2\right)\left(1 + \frac{(2\pi\alpha')^2B^2}{g_{xx}^2}-
\frac{\a^2/\N^2}{|g_{tt}|g_{\psi\psi} g_{xx}^3g_{ss}^3}\right)\\
&-& g_{xx} \left( \frac{\beta}{\N} + \frac{(2\pi\alpha')B\omega
r^4}{\rho^4}\right)^2\Bigg]^{1/2}.
\nonumber \eea %
$\beta=\frac{\delta \hat{S}_{D7}}{\delta A'_z}$ and double hat means that the second Legendre
transformation has been applied. Using \eqref{seq Legen} the equation of motion for $R(r)$ can be easily found and a similar analysis leading to \eqref{ahmad2} reduces to %
\bse\begin{align}
\label{worldvolumehorizonk}\frac{\gamma^2}{2}\frac{f(\rho_*)^2}{H(\rho_*)}& -
\frac{R_*^2 \omega^2}{\rho_*^4}\\ \nonumber
& -\frac{5k}{\rho^{20}_*H(\rho_*)^6}\left(\frac{ \gamma^2}{2}\frac{f(\rho_*)^2 U_1(\rho_*)}{H(\rho_*)}+\frac{3\omega^2 R_*^2 T(\rho_*)}{H(\rho_*)^6}\right)=0, \\
\label{ahmad3}\beta &=-(2\pi\alpha')\N B \omega \frac{ r_*^4 }{\rho_*^4}, \\
\label{ahmad4}\alpha &=-\frac{1}{4}r_*^3 \gamma ^4 R_* f(\rho_*)H(\rho_*)\sqrt{1+\frac{4(2\pi\alpha')^2 B^2\left(1-\frac{25 k T(\rho_*)}{\rho_*^{16}H(\rho_*)^6}\right)}{\gamma ^4 \rho_*^4H(\rho_*)^2 }}\\ \nonumber
&\times  \left(1+\frac{15 k\text{  }T(\rho_*)}{\rho_*^{16}H(\rho_*)^6}\right)^2 \sqrt{1-\frac{5 k\text{  }U_1(\rho_*)}{\rho_*^{20}H(\rho_*)^6}}~.
\end{align}\ese %

\end{document}